\documentclass[aps,pre,notitlepage,nobibnotes,nofootinbib,%
superscriptaddress,onecolumn,a4paper,10pt]{revtex4-1} 

\makeatletter
\def\p@subsection{}
\def\p@subsubsection{}
\makeatother

\setcitestyle{authoryear,round}
\usepackage[T1]{fontenc}
\usepackage[utf8]{inputenc}
\usepackage{microtype}
\usepackage{lmodern}
\usepackage[english]{babel}
\usepackage{mathtools}

\usepackage[hidelinks]{hyperref}
\usepackage{cleveref}
\usepackage{pgf}
\usepackage{enumerate}
\usepackage[low-sup]{subdepth}

\usepackage{bm}
\let\vec\bm%

\newcommand{\tr}{\operatorname{tr}}

\usepackage{mleftright}
\let\left\mleft%
\let\right\mright%

\newcommand{\pder}[2]{\frac{\partial#1}{\partial#2}}
\newcommand{\pdder}[2]{\frac{\partial^2#1}{\partial#2^2}}
\newcommand{\pader}[2]{\partial{#1}/\partial{#2}}

\newcommand{\phm}{\phantom{-}}
\newcommand{\Hessian}{\mathcal{Q}}
\newcommand{\cyl}{\rho}
\newcommand{\charge}{\varrho^\mathrm{s}}
\newcommand{\current}{\mathcal{J}^\mathrm{s}_\varphi}

\begin{document}

\title{Orbital stability in static axisymmetric fields}
\author{Gopakumar Mohandas}
\author{Tobias Heinemann}
\author{Martín E.~Pessah}
\affiliation{Niels Bohr International Academy, Niels Bohr Institute,
  Blegdamsvej 17, 2100 Copenhagen, Denmark}
\date{\today}

\begin{abstract}
  We investigate the stability of test-particle equilibrium orbits in
  axisymmetric, but otherwise arbitrary, gravitational and electromagnetic
  fields. We extend previous studies of this problem to include a toroidal
  magnetic field. We find that, even though the toroidal magnetic field does
  not alter the location of the circular orbits, it enters the problem as a
  gyroscopic force with the potential to provide gyroscopic stability. This is
  in essence similar to the situation encountered in the reduced three-body
  problem where rotation enables stability around the local maxima of the
  effective potential. Nevertheless, we show that gyroscopic stabilization by
  a toroidal magnetic field is impossible for axisymmetric force fields in
  source-free regions because in this case the effective potential does not
  possess any local maxima. As an example of an axisymmetric force field with
  sources, we consider the classical problem of a rotating, aligned
  magnetosphere. By analyzing the dynamics of halo and equatorial particle
  orbits we conclude that axisymmetric toroidal fields that are antisymmetric
  about the equator are unable to provide gyroscopic stabilization.  On the
  other hand, a toroidal magnetic field that  does not vanish at the equator
  can provide gyroscopic stabilization for positively charged particles in
  prograde equatorial orbits.
\end{abstract}

\maketitle

\section{Introduction}

The study of particle dynamics in axisymmetric fields is important in a
variety of problems spanning a wide range of scales in nature.  The axial
symmetry can be exploited to reduce the number of degrees of freedom. In the
reduced phase space, the scalar potential, which conservative forces derive
from, is replaced by an effective potential that includes the toroidal kinetic
energy. Critical points of the effective potential are equilibria of the
reduced system. These correspond to circular orbits in three-dimensional
space. \citet{Howard1999a} provides a thorough overview of the stability of
circular orbits in axisymmetric gravitational and electromagnetic fields. The
magnetic fields considered there are assumed to be purely poloidal. While this
is a reasonable assumption to make in many axisymmetric systems, it is not
difficult to think of examples where toroidal magnetic fields play a
significant role. Such examples include toroidal fusion devices and galactic
disks.

In this paper, we generalize \citet{Howard1999a} analysis addressing the
stability of circular orbits by including a toroidal axisymmetric magnetic
field. Axisymmetric poloidal and toroidal magnetic fields enter the dynamics
in a fundamentally different way. All the effects related to the poloidal
magnetic field can be encapsulated in the effective potential together with
the gravitational contribution, whereas this is never the case for the
toroidal part. As we detail below, this crucial difference makes it possible
for the toroidal magnetic field to provide stability in regions of parameter
space where a purely poloidal field  cannot. This is a manifestation of the
phenomenon known as gyroscopic stability, which is usually associated with the
Coriolis force.

The rest of the paper is organized as follows. In
\cref{sec:equations-of-motion} we state the equations of motion for a test
particle in reduced phase space by introducing the Routhian and an effective
potential that depends only on the poloidal flux function. In
\cref{sec:stability-of-circular-orbits} we analize the stability of circular
orbits and find the conditions for gyroscopic stabilization via a toroidal
magnetic field. In particular, we show that in source-free regions, gyroscopic
stabilization via a toroidal magnetic field is impossible. In
\cref{sec:magnetosphere} we illustrate some of the implications of our
findings by analysing the problem of a rotating magnetosphere. We conclude by
discussing our results in \cref{sec:discussion}.

\section{Equations of motion}\label{sec:equations-of-motion}

The motion of a classical, non-relativistic\footnote{This is the most general
  single particle Lagrangian compatible with Galilean invariance
\citep{Jauch1964,Roman1974}.} particle is governed by the Lagrangian per unit
particle mass
\begin{equation}
  \label{eq:lagrangian}
  L = \frac{1}{2}\dot{\vec{r}}^2 + \vec{A}\cdot\dot{\vec{r}} - \Phi,
\end{equation}
where $\dot{\vec{r}}$ denotes the time derivative of the particle's position
vector. We have absorbed the coupling constants (i.e.,\ charge and mass) in
the scalar potential $\Phi$ and the vector potential $\vec{A}$, both of which
are assumed time-independent. In the following, we will refer to $\Phi$ and
$\vec{A}$ as electromagnetic potentials. Note, however, that $\Phi$ can
include the gravitational potential and/or the centrifugal potential and
$\vec{A}$ can include a contribution accounting for the Coriolis force that
arises in a rotating frame. The equation of motion derived from
\cref{eq:lagrangian} via the Euler-Lagrange equation is
$\ddot{\vec{r}}=\vec{E}+\dot{\vec{r}}\times\vec{B}$, where the electric field
is $\vec{E}=-\nabla\Phi$ and the magnetic field is
$\vec{B}=\nabla\times\vec{A}$. We note that if only electromagnetic forces are
present, then $\vec{E}$ and $\vec{B}$ differ from the true electromagnetic
fields by a factor equal to the charge-to-mass ratio.

\subsection{Motion in reduced phase space}

We work in cylindrical coordinates $(\cyl,\varphi,z)$ and assume that the
system is symmetric about the $z$-axis. This means that neither $\vec{A}$ nor
$\Phi$ depend on the cyclic coordinate $\varphi$. From the Euler-Lagrange
equation $d/dt(\pader{L}{\dot{\varphi}})=\pader{L}{\varphi}$ it then follows
that the generalized angular momentum $p_\varphi=\pader{L}{\dot{\varphi}}$ is
an integral of motion. Substituting \cref{eq:lagrangian} we obtain
\begin{equation}
  \label{eq:angular-momentum}
  p_\varphi = \cyl^2\dot{\varphi} + \psi,
\end{equation}
where we have introduced the poloidal flux function
\begin{equation}
  \label{eq:poloidal-flux}
  \psi = \cyl A_\varphi,
\end{equation}
in terms of which the magnetic field is given by
$\vec{B}=\nabla\psi\times\nabla\varphi+\cyl B_\varphi\nabla\varphi$.

Since $p_\varphi$ is an integral of motion, the dimensionality of the problem
may be reduced by one. For this we introduce the Routhian
$R=L-\omega p_\varphi$, where
\begin{equation}
  \label{eq:angular-velocity}
  \omega = \frac{1}{\cyl^2}(p_\varphi - \psi)
\end{equation}
is equal to the angular velocity $\dot{\varphi}$ expressed through
\cref{eq:angular-momentum} as a function of $\cyl$ and $z$. With
\cref{eq:angular-velocity} the Routhian is given by
\begin{equation}
  \label{eq:routhian}
  R = \frac{1}{2}(\dot{\cyl}^2 + \dot{z}^2)
  + A_\cyl\dot{\cyl} + A_z\dot{z} - U,
\end{equation}
where
\begin{equation}
  \label{eq:effective-potential}
  U = \Phi + \frac{\cyl^2\omega^2}{2}
\end{equation}
is the \emph{effective} potential.

The equations of motion of the reduced system are
$d/dt(\pader{R}{\dot{q}^i})=\pader{R}{q^i}$ for $q^i=(\cyl,z)$ or
\begin{equation}
  \label{eq:motion}
  \begin{aligned}
    \ddot{\cyl} + B_\varphi\dot{z} + \pader{U}{\cyl} &= 0, \\
    \ddot{z} - B_\varphi\dot{\cyl} + \pader{U}{z} &= 0.
  \end{aligned}
\end{equation}
These equations possess the energy integral
\begin{equation}
  \label{eq:hamiltonian}
  H = \frac{1}{2}\left(\dot{\cyl}^2 + \dot{z}^2\right) + U,
\end{equation}
which evidently is independent of the \emph{toroidal} magnetic field. This is
because the force due to the toroidal magnetic field does not do work in the
reduced configuration space. Such forces are called gyroscopic forces
\citep{Thomson1883,Thomson1883a}. Very much in contrast to this, the force due
to the poloidal magnetic field, which in fact is gyroscopic before reduction,
has become a potential force in the reduced configuration space. A reduced
system of the form \labelcref{eq:motion} with $B_\varphi\ne0$ is said to be
gyroscopically constrained or coupled
\citep[see e.g.][]{Rumiantsev1966,Merkin1996}.

\subsection{Hamiltonian formalism}

Before moving on to study the stability of equilibrium solutions of the
equations of motion, we note that the reduced system may also be described
using a Hamiltonian formalism. In a gyroscopically coupled system this is best
done by working in non-canonical phase space coordinates
\begin{equation}
  \label{eq:phase-space-coordinates}
  w^\alpha = (\cyl, z, \dot{\cyl}, \dot{z}),
\end{equation}
see e.g.\ \citet{Littlejohn1979,Littlejohn1982} or \citet{Bolotin1995}. In
these coordinates, the equations of motion are
\begin{equation}
  \label{eq:hamilton}
  \dot{w}^\alpha = J^{\alpha\beta}\pder{H}{w^\beta},
\end{equation}
where the Hamiltonian $H$ is defined in \cref{eq:hamiltonian} and the Poisson
matrix $J^{\alpha\beta}$ is given by
\begin{equation}
  \label{eq:poisson-matrix}
  J^{\alpha\beta} = \begin{pmatrix}
    \phm 0 & \phm 0 & 1 & \phm 0 \\
    \phm 0 & \phm 0 & 0 & \phm 1 \\
    -1 & \phm 0 & 0 & -B_\varphi \\
    \phm 0 & -1 & B_\varphi & \phm 0
  \end{pmatrix}.
\end{equation}
It is straightforward to verify that given the definitions in
\cref{eq:hamilton,eq:poisson-matrix}, Hamilton's equations in the form of
\cref{eq:hamilton} are equivalent to the equations of motion
\labelcref{eq:motion}. We also note that the Poisson bracket defined by
\mbox{$\{f,g\}=(\pader{f}{w^\alpha})J^{\alpha\beta}(\pader{g}{w^\beta})$}
satisfies the Jacobi identity
$\{f,\{g,h\}\!\}+\{g,\{h,f\}\!\}+\{h,\{f,g\}\!\}=0$ for any
$B_\varphi(\cyl,z)$, as it should.

\section{Stability of circular orbits}\label{sec:stability-of-circular-orbits}

In this section we discuss the stability of equilibrium solutions to the
equations of motion \labelcref{eq:motion}. We note that all that is said here
in fact holds for arbitrary forms of the effective potential. This means that
for instance the restricted three-body problem is within the scope of our
discussion. Only in \cref{sec:magnetosphere} will we specialize to effective
potentials of the form given in \cref{eq:effective-potential}, in which the
scalar potential $\Phi$ is independent of the canonical angular momentum
$p_\varphi$.

\subsection{Stability criteria}

Equilibria of the reduced system described by \cref{eq:routhian} are solutions
with $\dot{\cyl}=\dot{z}=0$. The angular velocity $\dot{\varphi}$ is, however,
in general non-zero. Equilibria of the reduced system thus correspond to
uniformly rotating solutions of the original system. In the classic literature
\citep[e.g.][]{Routh1884}, such solutions are known as steady motions. A more
modern term is relative equilibria \citep[e.g.][]{Marsden1974}.\footnote{It
  should be noted that relative equilibria potentially encompass a much wider
  class of solutions than just steady motions. Like steady motions, relative
  equilibria are obtained by reduction through symmetry, but unlike steady
  motions, relative equilibria allow for the underlying symmetry group to be
non-Abelian.} In the following we will refer to these solutions simply as
\emph{circular orbits}.

Inspection of the equations of motion \labelcref{eq:motion} reveals that
circular orbits are stationary points of the effective potential, i.e.\ points
at which $\delta U=0$ for arbitrary variations $\delta\cyl$ and $\delta z$.
Their location is evidently independent of the toroidal magnetic field
$B_\varphi$. This agrees with the expectation that because the magnetic force
is perpendicular to the velocity, the toroidal magnetic field should of course
have no effect on strictly circular orbits.

Whether or not circular orbits can be expected to actually occur in nature
(along with nearly circular orbits in their vicinity) depends on their
stability. Various notions of stability exist in the literature
\citep[see e.g.][]{Holm1985}. Arguably the most important one for practical
purposes is due to Lyapunov: an equilibrium $w^\alpha_\mathrm{e}$, with
$w^\alpha$ defined in \cref{eq:phase-space-coordinates}, is stable for every
$\varepsilon>0$ if there is a $\delta>0$ such that if
$|w^\alpha(0)-w^\alpha_\mathrm{e}|<\delta$ then
$|w^\alpha(t)-w^\alpha_\mathrm{e}|<\varepsilon$ for $t>0$. It is important to
note that $w^\alpha(t)$ in this definition evolves according to the nonlinear
equations of motion \labelcref{eq:hamilton}. By contrast, spectral stability
is concerned with the spectrum of the Hamiltonian matrix
$J^{\alpha\beta}\partial^2H/\partial w^\beta\partial w^\gamma$, obtained by
linearizing \cref{eq:hamilton}. A Hamiltonian system is spectrally stable if
all eigenvalues of this matrix lie on the imaginary axis. Lyapunov stability
implies spectral stability but not vice versa \citep{Holm1985}. In the
following, stability will generally be synonymous with Lyapunov stability ---
the stronger of the two notions --- unless specified otherwise.

The discussion in the above paragraph has an important caveat: the canonical
angular momentum $p_\varphi$ (the cyclic integral) is assumed fixed. Truly
arbitrary perturbations would also allow $p_\varphi$ to be varied. It is often
argued that in practice the restriction of fixed cyclic integrals is
unimportant because, as \citet{Pars1965} writes, ``if we do allow small
changes in the [cyclic integrals] we are merely transferring our attention to
oscillations about a neighboring state of steady motion.'' This argument is
originally due to Lyapunov \citep[see][]{Rumiantsev1966}. It may of course be
that there is no neighboring state of steady motion, in which case this
argument fails and a more elaborate approach becomes necessary
\citep{Salvadori1953}. A comprehensive discussion is given by
\citet{Hagedorn1971}. In the following we will ignore these subtleties and
take $p_\varphi$ as fixed.

According to Routh's theorem \citep{Routh1884}, a circular orbit is stable if
it corresponds to an isolated minimum of the effective potential. This is the
case if the second variation $\delta^2 U>0$ for any non-zero $\delta\cyl$
and/or $\delta z$. Expressed in terms of the Hessian $\Hessian$ of the
effective potential, whose trace and determinant are given by
\begin{equation}
  \tr\Hessian =
  \frac{\partial^2 U}{\partial \cyl^2}
  + \frac{\partial^2 U}{\partial z^2}
  \quad\textrm{and}\quad
  \det\Hessian =
  \frac{\partial^2 U}{\partial\cyl^2} \frac{\partial^2 U}{\partial z^2} -
  {\left(\frac{\partial^2 U}{\partial\cyl\partial z}\right)}^2,
\label{eq:treace-determinant}
\end{equation}
a critical point of the effective potential is an isolated minimum if and only
if
\begin{equation}
  \label{eq:local-minimum}
  \tr\Hessian > 0
  \quad\textrm{and}\quad
  \det\Hessian > 0.
\end{equation}
If the inequalities in \cref{eq:local-minimum} are satisfied, then the total
energy at equilibrium is positive definite (i.e.\ $\delta^2 H>0$ for any
non-zero $\delta w^\alpha$) and can thus be used as a Lyapunov function to
prove Routh's theorem using Lyapunov's direct method \citep{Merkin1996}. In
the absence of a toroidal magnetic field ($B_\varphi=0$), in which case the
system is gyroscopically decoupled, the converse is also true
\citep{Lyapunov1907,Malkin1959,Chetaev1961,Hagedorn1971,Rumyantsev1993}:
circular orbits are unstable if they do \emph{not} minimize $U$ locally.

\subsection{Gyroscopic stabilization}

In the presence of a toroidal magnetic field, the system is gyroscopically
coupled. In this case, all circular orbits located at isolated minima of $U$
are still stable. However, there may now also exist stable circular orbits
located at isolated \emph{maxima} of the effective potential, where
$\delta^2U<0$. This is known as gyroscopic stabilization
\citep{Thomson1883,Thomson1883a,Chetaev1961,Merkin1996}. All orbits are
spectrally unstable at \emph{saddle points} ($\det\Hessian<0$).

A note about energetics is in order here. Isolated minima of the effective
potential correspond to isolated minima of the total energy because the
kinetic energy is positive definite. In other words $\delta^2H>0$.
Definiteness of $\delta^2H$ is referred to as formal or energetic stability
\citep{Holm1985,Scheeres2006}. It is a sufficient but not necessary condition
for Lyapunov stability. The toroidal magnetic field does not affect the
energetic stability of circular orbits. It can, however, stabilize
energetically unstable orbits, namely isolated maxima of $U$, for which
$\delta^2H$ is indefinite because $\delta^2U<0$.

\begin{figure}
  \centerline{\includegraphics{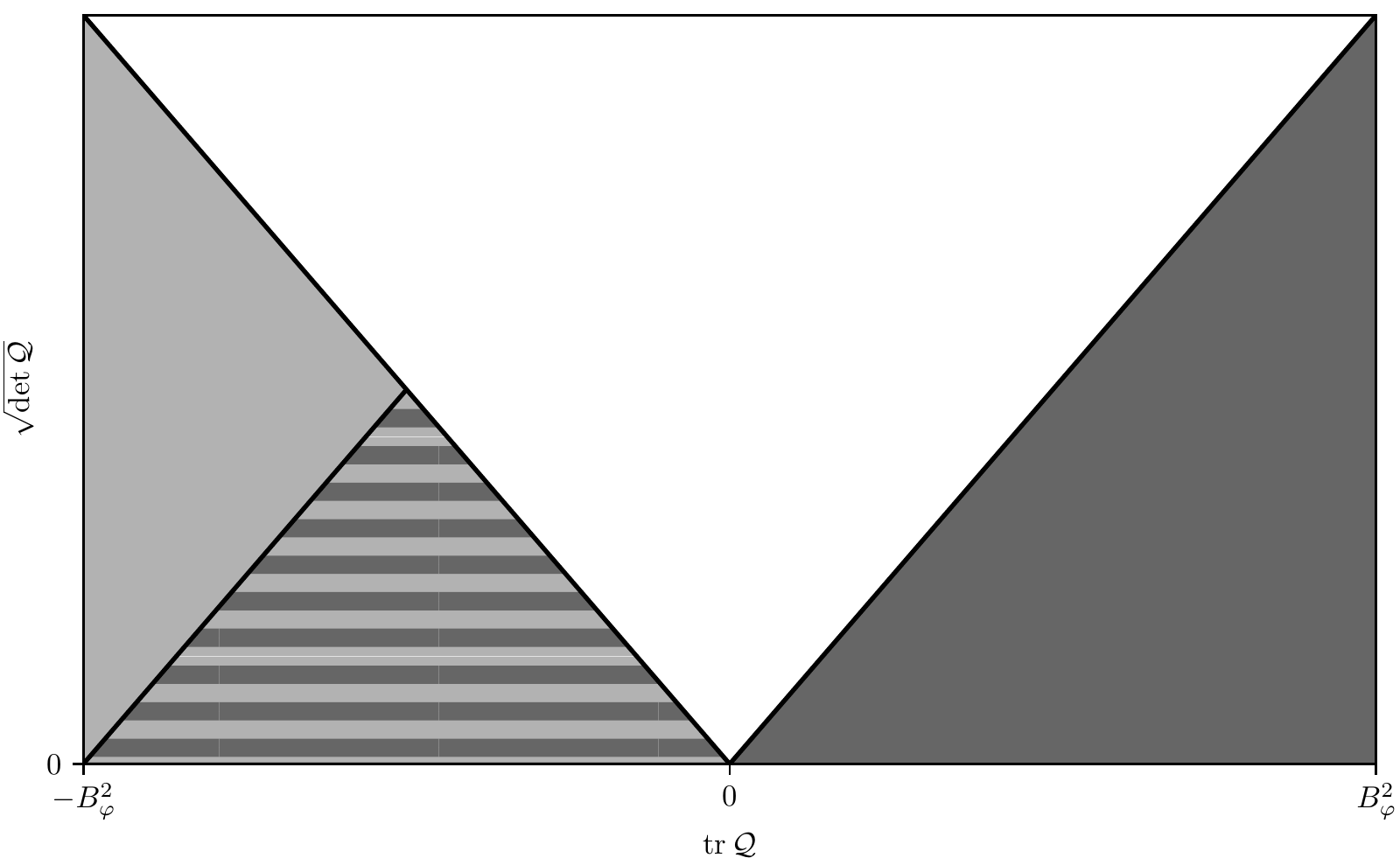}}
  \caption{%
    Orbits in the dark-grey region are formally stable.  Orbits in the
    stripped region are gyroscopically stabilized by a toroidal magnetic field
    $B_\varphi$. Orbits in the light-grey region outside the stripped region
    are unstable. The fraction of this parameter space that is actually
    populated with orbits depends on the specific structure of the
    electromagnetic potentials. Note that the trace
    and the determinant of a real symmetric $2\times 2$-matrix satisfy
    $2\sqrt{\det\Hessian} \le {|\tr\Hessian|}$, thus the white area is
    inaccessible.
  }\label{fig:stabilization}
\end{figure}

In order to determine the conditions for gyroscopic stabilization to occur, we
first carry out a spectral stability analysis. Considering infinitesimal
perturbations of the linearized equations of motion \cref{eq:motion} leads to
the characteristic polynomial
\begin{equation}
  \label{eq:dispersion-relation}
  \sigma^4 + (B_\varphi^2 + \tr\Hessian)\sigma^2 + \det\Hessian = 0.
\end{equation}
The roots of this equation comprise the spectrum of the Hamiltonian matrix
$J^{\alpha\beta}\partial^2 H/\partial w^\beta\partial w^\gamma$ mentioned
above. A comprehensive discussion of \cref{eq:dispersion-relation} is given in
\citet{Bloch1994}, see also \citet{Chetaev1961} and \citet{Haller1992}.
Isolated maxima of the effective potential are critical points where $\delta^2
U<0$ or, equivalently, $\det\Hessian>0$ and $\tr\Hessian<0$. Depending on the
strength of the toroidal field, only a subset of these maxima are
gyroscopically stabilized. The precise conditions are
\begin{equation}
  \label{eq:gyro-stability}
  -B_\varphi^2 < \tr\Hessian < 0
  \quad\textrm{and}\quad
  0<4\det\Hessian < {\left(\tr\Hessian + B_\varphi^2\right)}^2.
\end{equation}
Note that for a given maximum of the effective potential, it is always
possible to satisfy these inequalities for large enough $B_\varphi$. A visual
representation of the inequalities in
\cref{eq:local-minimum,eq:gyro-stability} is given in
\cref{fig:stabilization}.

If the inequalities in \cref{eq:gyro-stability} are satisfied, then the
circular orbit is gyroscopically stabilized in the spectral sense: all
eigenvalues $\sigma$ as given by the roots of \cref{eq:dispersion-relation}
lie on the imaginary axis. In order to show that circular orbits can be
gyroscopically stabilized in the Lyapunov sense is more challenging. Since $U$
is negative definite at an isolated maximum, the total energy $H$ defined in
\cref{eq:hamiltonian} is indefinite and thus cannot be used as a Lyapunov
function to prove stability using the direct method.

Instead, Lyapunov stability at isolated maxima of the effective potential can
be demonstrated with the help of the Kolmogorov-Arnold-Moser (KAM) theorem. At
a spectrally stable isolated maximum, where eqs.~\labelcref{eq:gyro-stability}
are satisfied, the eigenvalues $\pm\sigma_1$ and $\pm\sigma_2$ of the
linearized system, given by the roots of the characteristic polynomial
\labelcref{eq:dispersion-relation}, are purely imaginary. If these eigenvalues
are non-resonant ($\sigma_1^2\ne n^2\sigma_2^2$ for $n=1,2,3$), then the
nonlinear dynamics of the system close to equilibrium is nearly integrable.
The integrable, linear dynamics takes place on two-dimensional tori in phase
space and the KAM theorem ensures these tori persist under nonlinear
perturbations, provided certain non-degeneracy conditions are met
\citep{Arnold1963,Haller1992}. Note that since $p_\varphi$ is assumed fixed,
these considerations strictly speaking only demonstrate \emph{conditional}
Lyapunov stability for the reduced system.

It is important to stress that both resonance and degeneracy occur with
probability zero in continuous (real valued) parameter space. Moreover, if the
system is resonant it may still be stable \citep{Sokolskii1974}. Likewise,
non-degeneracy is sufficient but not necessary for stability: in systems with
two degrees of freedom (such as the present one) degenerate equilibria are, as
a rule, stable \citep[sec.~6.3.6.B]{Arnold2006}.

\subsection{The effects of dissipation}

The above considerations need to be amended if the system is dissipative.
Minima of the effective potential remain stable when dissipation is added to
the system, however, all other equilibria are unstable no matter how small
(but finite) the dissipation is. In particular, gyroscopically stabilized
equilibria, which correspond to maxima of the effective potential, lose their
stability if dissipation is added
\citep{Thomson1883,Thomson1883a,Chetaev1961,Rumiantsev1966,Haller1992}. Loss
of gyroscopic stabilization due to dissipation is an instance of a wider class
of phenomena known as dissipation-induced instabilities
\citep{Bloch1994,Krechetnikov2007}.

In practice gyrcosopic stabilization is thus only a transient phenomenon. The
growth rate of these instabilities, which are generally proportional to the
dissipation rate \citep{MacKay1991,Bloch1994}, is much smaller than the growth
rates in the absence of gyroscopic forces. \citet{Thomson1883,Thomson1883a}
refer to this as temporary, as opposed to secular, stability.

The fact that in realistic systems, gyroscopic forces are not able to truly
stabilize an otherwise unstable equilibrium does not diminish the significance
of gyroscopic stabilization by very much. This can be seen in the restricted
three-body problem applied to the Sun-Jupiter system. The critical points of
the effective potential in this problem are either saddle points ($L_1$,
$L_2$, and $L_3$) or local maxima ($L_4$ and $L_5$). Gyroscopic stability at
$L_4$ and $L_5$ is provided by the Coriolis force. The Trojan asteroids are
are found to cluster around $L_4$ and $L_5$ even though they are subject to a
dissipative force due to nebular drag and hence to dissipation induced
instability.

\subsection{Impossibility of gyroscopic stabilization in source-free regions}

In the previous section we have seen that whether gyroscopic stabilization is
possible depends on the sign of $\tr\Hessian$. In order to calculate the
trace Hessian, we need to compute its diagonal elements. These are given by
\begin{equation}
  \label{eq:D2U}
  \pdder{U}{\cyl} = \pdder{\Phi}{\cyl} - \omega\pdder{\psi}{\cyl}
  + 3\omega^2 + 4\omega B_z + B_z^2
  \quad\textrm{and}\quad
  \pdder{U}{z} =
  \pdder{\Phi}{z} - \omega\pdder{\psi}{z} + B_\cyl^2,
\end{equation}
where the poloidal magnetic field components are
$B_\cyl=-\cyl^{-1}\pader{\psi}{z}$ and $B_z=\cyl^{-1}\pader{\psi}{\cyl}$.
Adding up \cref{eq:D2U} yields the trace of the Hessian. The resulting
expression can be simplified with the help of Gauss' law
$\nabla\cdot\vec{E}=\charge$, i.e.
\begin{equation}
  \label{eq:gauss}
  \pdder{\Phi}{\cyl} + \frac{1}{\cyl}\pder{\Phi}{\cyl} + \pdder{\Phi}{z}
  = - \charge
\end{equation}
and Ampère's law $\nabla\times\vec{B}=\vec{\mathcal{J}}^\mathrm{s}$, whose
toroidal component is given by
\begin{equation}
  \label{eq:ampere}
  \pdder{\psi}{\cyl} - \frac{1}{\cyl}\pder{\psi}{\cyl} + \pdder{\psi}{z} =
  -\cyl\current.
\end{equation}
In the plasma physics literature, the differential operator acting on $\psi$
in \cref{eq:ampere} is known as the Grad-Shafranov operator
\citep[see e.g.][]{Almaguer1988}. With \cref{eq:gauss,eq:ampere}, evaluating
the trace at equilibrium yields
\begin{equation}
  \label{eq:trace}
  \tr\Hessian =
  \omega^2 + {(\omega + B_z)}^2 + B_\cyl^2 - \charge + \cyl\omega\current,
\end{equation}
where we have used
\begin{equation}
  \label{eq:DU}
  \pder{\Phi}{\cyl} - \omega\pder{\psi}{\cyl} = \cyl\omega^2
  \quad\textrm{and}\quad
  \pder{\Phi}{z} - \omega\pder{\psi}{z} = 0.
\end{equation}
The first three terms on the right hand side of \cref{eq:trace} are all
non-negative. Thus, in source-free regions, where $\charge=0$ and
$\current=0$, the effective potential has no local maxima and gyroscopic
stabilisation cannot occur. We note that matter distributions make a
\emph{negative} contribution to $\charge$ and hence a positive contribution to
$\tr\Hessian$. We also note that poloidal currents, i.e.\ sources of the
toroidal magnetic field, do not enter \cref{eq:trace} at all.

In addition to physical sources in the form or matter and charge distributions
there may also be fictitious sources that arise in a rotating frame. For
instance, associated with the centrifugal force is a fictitious charge
distribution $\charge=2\Omega^2$. That being said, in
\cref{sec:rotating-frame} we show that the stability of circular orbits is not
affected by a transformation to a rotating frame of reference.

\section{Motion in a rotating magnetosphere}\label{sec:magnetosphere}

As an example of astrophysical relevance where gyroscopic stability by a
toroidal magnetic field is possible we analyse the stability of circular
orbits in the classical problem of a rotating magnetosphere.  We consider the
same model studied by \citet{Howard1999,Howard2000} and \citet{Dullin2002},
where the poloidal magnetic field and the planetary gravitational potential
are due to a point dipole and a point mass, respectively.

The poloidal flux function is given by
\begin{equation}
  \label{eq:poloidal-flux-function}
  \psi = \frac{\gamma\cyl^2}{r^3},
\end{equation}
where $r$ is the spherical radius. The parameter $\gamma$ is proportional to
the product of the charge to mass ratio and the dipole strength. Without loss
of generality we take the magnetic dipole to point along the $z$-axis. With
this convention, $\gamma >0$ for positive charges.  The poloidal magnetic
field components $B_\cyl=-\cyl^{-1}\pader{\psi}{z}$ and
$B_z=\cyl^{-1}\pader{\psi}{\cyl}$ derived from
\cref{eq:poloidal-flux-function} are
\begin{equation}
  B_\cyl = \frac{3\gamma\cyl z}{r^5}
  \quad\textrm{and}\quad
  B_z = \frac{\gamma(2z^2 - \cyl^2)}{r^5}.
\end{equation}
The scalar potential is
\begin{equation}
  \label{eq:scalar-potential}
  \Phi = - \frac{\mu}{r} + \Omega\psi.
\end{equation}
The first term is the gravitational potential. The second term arises from the
requirement that the electric field vanishes in a frame rotating with the
planetary rotation rate $\Omega$.

The current density that derives from \cref{eq:poloidal-flux-function} via
Ampère's law \labelcref{eq:ampere} vanishes away from the origin. However, the
charge density that derives from \cref{eq:scalar-potential} via Gauss' law
\labelcref{eq:gauss} does not vanish and is given by
\begin{equation}
  \label{eq:corotational-charge}
  \charge = -2\Omega B_z
\end{equation}
for $r>0$. This charge density, known in the astrophysical literature as the
Goldreich-Julian charge density \citep{Goldreich1969}, is distributed
continuously throughout space. Because of this, positivity of the trace in
\cref{eq:trace} is no longer ensured.

In order to assess what orbits can be subject to gyroscopic stabilisation, we
analyse the characteristics of equilibria, focusing our attention on the
regions of parameter space associated with negative $\tr\Hessian$. The
calculations involved in obtaining $\det\Hessian$ and $\tr\Hessian$ in
\cref{eq:treace-determinant} have already been carried out in
\citet{Dullin2002}.  For the reader's convenience, and because we use
different notation, we restate the results that are relevant to the discussion
here.

Using \cref{eq:poloidal-flux-function} and \cref{eq:scalar-potential} we
obtain the location of the equilibrium circular orbits by requiring that  the
first variation of the effective potential $U$ in
\cref{eq:effective-potential} vanishes, i.e.\ by requiring
\begin{align}
  \label{eq:effective-potential-partial-rho}
  \pder{U}{\cyl} &= \frac{\cyl}{r^5}\bigg[
    \mu r^2 + 3\gamma\cyl^2(\omega - \Omega)
    - 2\gamma(\omega - \Omega)r^2 - \omega^2 r^5
  \bigg]=0, \\
  \pder{U}{z} &= \frac{z}{r^5}\bigg[
    \mu r^2 + 3\gamma\cyl^2(\omega - \Omega)
  \bigg]=0.
  \label{eq:effective-potential-partial-z}
\end{align}
Setting to zero each of the two factors on
\cref{eq:effective-potential-partial-z} leads to equatorial and halo orbits
respectively. We analyse each of these cases separately.

\subsection{Equatorial orbits}

\begin{figure}
  \centerline{\includegraphics{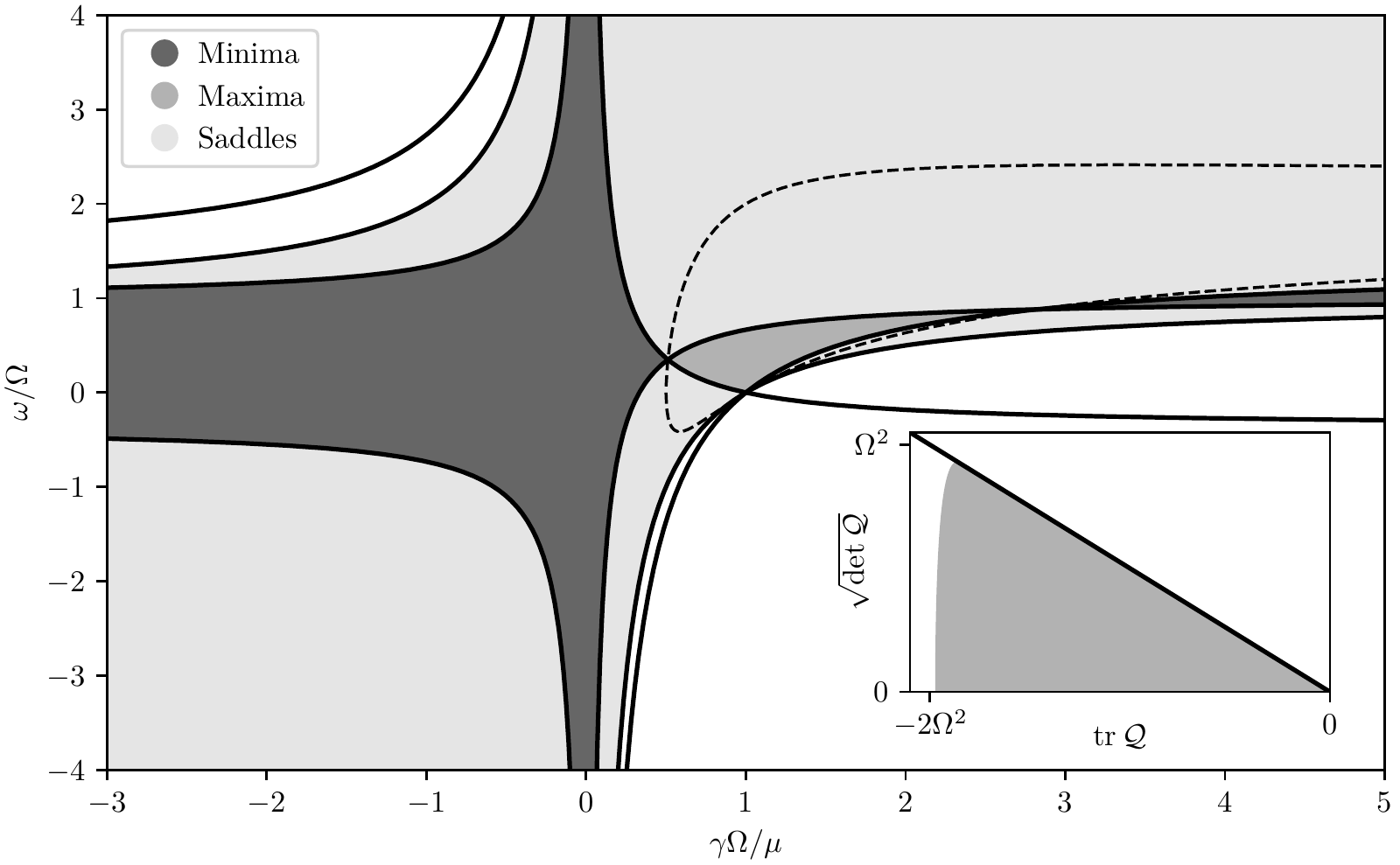}}%
  \caption{Stability diagram for equatorial orbits. The region where
    equatorial orbits exist are shaded light gray. Regions of formal stability
    are shaded dark gray. The only difference with respect to
    \citet{Dullin2002} is that we also indicate (by the intermediate shade of
    gray) the region where gyroscopic stabilization can in principle occur.
    This figure is consistent with \cref{eq:trace,eq:corotational-charge} in
    that, at the equator, $B_z<0$ for $\gamma>0$ and thus the term $+2\Omega
    B_z$ makes a negative contribution to the trace. Note that $B_z<0$ at the
    equator corresponds to a rotating dipole for which the magnetic moment and
    the angular velocity are aligned. The inset shows the region spanned by
    the orbits for which gyroscopic stabilization is possible in the parameter
    space defined in \cref{fig:stabilization}. Note that a toroidal magnetic
    field $B_\varphi\ge\sqrt{2}\Omega$ provides gyroscopic stabilization for
    all the orbits for which this is possible.}%
  \label{fig:dullin}
\end{figure}

Equatorial orbits lie within the plane $z=0$. Their radial location is
obtained from \cref{eq:effective-potential-partial-rho} with $r=\cyl$. The
result is
\begin{equation}
  \label{eq:equator-coordinates}
  \cyl^3 = \frac{\mu + \gamma(\omega - \Omega)}{\omega^2}.
\end{equation}
The determinant and trace of the Hessian evaluated at equatorial equilibria
are given by
\begin{equation}
  \label{eq:determinant-equator}
  \det\Hessian =
  -\frac{1}{2\omega^2\cyl^9}
  \Big[3\gamma\left(\omega - \Omega\right) + \mu\Big]
  \Big[
    {(2\omega\gamma - \Omega\gamma + \mu)}^2 - 3{(\mu - \gamma\Omega)}^2
  \Big]
\end{equation}
and
\begin{equation}
  \label{eq:trace-equator}
  \tr\Hessian = \frac{1}{\omega^2\cyl^6}
  \left[
    \gamma^{2}{\left(\omega - 2\Omega\right)}^{2}
    + 2\gamma\mu\left(\omega - 3\Omega\right) + 2\mu^{2}
  \right].
\end{equation}
From these expressions we can easily compute the zeros of both $\det\Hessian$
and $\tr\Hessian$ and characterise the various regions of parameter space
according to their stability properties. The results are illustrated in
\cref{fig:dullin}, which is almost identical to Figure~7\ in
\citet{Dullin2002}, except that we also plot the curve along which the trace
vanishes.

In addition to the two energetically stable regions identified already by
\citet{Dullin2002}, there is now a region spanned by orbits corresponding to
particles with positive charge ($\gamma>0$) in prograde rotation ($\omega>0$).
These orbits can be stabilized via gyroscopic effects provided a sufficiently
strong toroidal field is present. We remark, however, that this is unlikely to
be the case as the toroidal field is typically anti-symmetric about the
equator in realistic magnetospheres \citep[see e.g.][]{Bunce2001}.

\subsection{Halo orbits}

The coordinates for circular halo orbits are obtained by setting to zero the
second term in \cref{eq:effective-potential-partial-z} and using this result
in \cref{eq:effective-potential-partial-rho}. This leads to
\begin{equation}
  \label{eq:halo-coordinates}
  r^3 = -\frac{2\gamma(\omega - \Omega)}{\omega^{2}}
  \quad\textrm{and}\quad
  \sin^2\vartheta = -\frac{\mu}{3\gamma\left(\omega - \Omega\right)},
\end{equation}
where $\vartheta$ is the angle subtended between the radius vector $\vec{r}$
and the $z$-direction such that $\cyl = r\sin\vartheta$ and
$z=r\cos\vartheta$.

The determinant and trace of the Hessian evaluated at halo equilibria are
given by
\begin{equation}
  \label{eq:determinant-halo}
  \det\Hessian =
  - \frac{16\gamma^{2}\mu}{3\omega^{4} r^{12}}
  \big[\omega^{2} - 4\Omega\omega + \Omega^{2}\big]
  \big[3\gamma\left(\omega - \Omega\right) + \mu\big]
\end{equation}
and
\begin{equation}
  \label{eq:trace-halo}
  \tr\Hessian = \frac{2\gamma^{2}}{\omega^{4} r^{9}}
  \left[
    \mu\left(3\omega^{2} - 4\Omega^{2}\right)
    - 4\gamma\left(\omega - \Omega\right)
    {\left(\omega - 2\Omega\right)}^{2}
  \right].
\end{equation}

\begin{figure}
  \centerline{\includegraphics{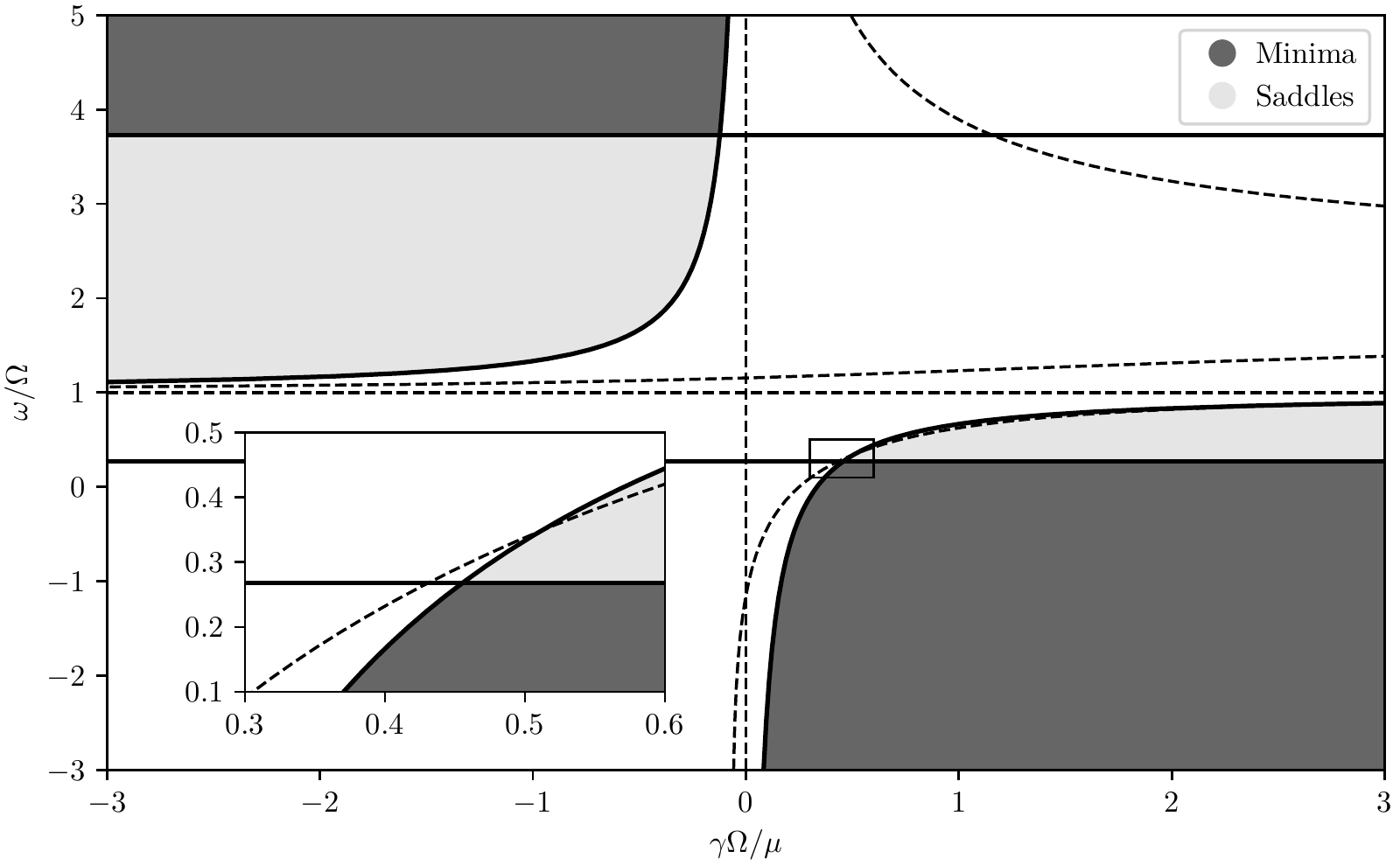}}%
  \caption{Stability diagram for halo orbits. This is analogous to Figure~9\
    in \citet{Dullin2002}. The dark regions are stable with $\det\Hessian>0$
    and $\tr\Hessian>0$. The sign of the determinant changes across solid
    lines and that of the trace across dashed lines. The solid hyperbola
    $\omega=\Omega-\mu/(3\gamma)$ is also the boundary of the region where
    circular halo orbits exist. Close inspection of the figure --- especially
    the inset --- reveals that there is a small region where (a) halo orbits
    exist and (b) the trace is negative. But the determinant is also negative
    in that region. The solid horizontal lines $\omega/\Omega=2\pm\sqrt{3}$
    thus mark the transition from local minima to saddle points. Gyroscopic
    stabilization is impossible because no halo orbits exist that correspond
    to local maxima of the effective potential.
  }\label{fig:halo}
\end{figure}

The regions of stability that derive from determining the signs of both
$\det\Hessian$ and $\tr\Hessian$ are illustrated in \cref{fig:halo}. This
figure is similar to Figure~9\ in \citet{Dullin2002}, except that we also plot
the curve along which the trace vanishes. Close inspection of this figure, see
in particular the inset, reveal that there are no regions where
$\det\Hessian>0$ and $\tr\Hessian<0$. We thus conclude that gyroscopic
stabilization of halo orbits via a toroidal magnetic field is impossible in an
aligned dipolar magnetosphere.

\section{Summary and Discussion}\label{sec:discussion}

\citet{Howard1999a} presented an overview of the stability of circular orbits
in combined axisymmetric gravitational and poloidal magnetic fields. In this
case, orbital stability is completely determined by the effective potential,
which contains all the dynamical effects arising from the magnetic field.
Under these assumptions, because the kinetic energy is positive definite,
equilibria are stable if and only if they minimize the effective potential.

In this paper, we have generalized this problem by including an axisymmetric
toroidal magnetic field, which cannot be included in the effective potential.
We pointed out that, unlike the poloidal field component, the toroidal
magnetic field does no work in the reduced phase space, i.e., the magnetic
force associated with it is gyroscopic\footnote{We point out that forces that
are gyroscopic in three-dimensional space do not necessarily act as gyroscopic
forces in the reduced space. This is indeed the case for a purely poloidal
axisymmetric magnetic field. We also note the Lagrangian can acquire
gyroscopic terms in the reduced space even in the absence of gyroscopic forces
in three-dimensional space.}. Thus even thought the toroidal field does not
influence the location of the circular orbits it can alter their stability
properties by enabling gyroscopic stability.

Absent dissipation, we carried out a spectral stability analysis and
determined the conditions for gyroscopic stabilization by an axisymmetric
toroidal magnetic field and summarized our results in
\cref{fig:stabilization}. We showed that, given a circular orbit in an
isolated local maxima of the effective potential, it is always possible to
find a sufficiently strong axisymmetric toroidal magnetic field to
gyroscopically stabilise it. Making use of the KAM theorem we concluded that
gyroscopic stability holds in the Lyapunov sense. In real systems, dissipative
processes prevent gyroscopic stability from being truly realized.
Nevertheless, this type of dissipation induced instabilities evolve on
timescales that are much longer than the growth rates of instabilities that
would operate in the absence of gyroscopic stabilization. We thus argue that
gyroscopic stabilization should be relevant.

We showed that the effective potential associated with combined axisymmetric
gravitational and poloidal magnetic fields does not present isolated local
maxima in source free regions, thus implying that gyroscopic stabilisation is
impossible in this case. As an example where sources are present, we
considered a rotating, aligned magnetosphere and investigated the effects of a
toroidal magnetic field. This is a generalization of the problem investigated
by Howard, who provided a detailed account of the stability properties of
equatorial \citep{Howard1999} and halo \citep{Howard2000} orbits of charged
dust-particles for an aligned, rotating dipole.  We found that there are no
equilibrium halo orbits that can be subject to gyroscopic stabilisation. We
also found, however, that there do exist prograde equatorial orbits for
positive charges for which a toroidal magnetic field that does not vanish at
the equator provides gyroscopic stabilization.

\appendix

\section{Circular orbits in a rotating frame}\label{sec:rotating-frame}

Let us carry out a coordinate transformation to a frame rotating with
a constant frequency $\Omega$ around the $z$-axis. The angular velocity in the
rotating frame is
\begin{equation}
  \label{eq:rotating-angular-velocity}
  \omega' = \omega - \Omega.
\end{equation}
The transformed scalar potential $\Phi'$ and the poloidal flux function
$\psi'$ are respectively given by
\begin{equation}
  \label{eq:rotating-scalar-potential}
  \Phi' = \Phi - \Omega\psi - \frac{1}{2}\cyl^2\Omega^2,
\end{equation}
and
\begin{equation}
  \label{eq:rotating-flux-function}
  \psi' = \psi + \cyl^2\Omega.
\end{equation}
The second term on the right hand side of \cref{eq:rotating-scalar-potential}
arises because the electrostatic potential transforms as the temporal
component of an ultra space-like four-vector
($A_0'=A_0-\vec{v}\cdot\vec{A}/c$). The last terms in
\cref{eq:rotating-scalar-potential,eq:rotating-flux-function} account for the
centrifugal force and the Coriolis force, respectively.

From \cref{eq:rotating-flux-function} it follows that the generalized angular
momentum, defined in \cref{eq:angular-momentum}, is invariant, i.e.
\begin{equation}
  \label{eq:rotating-angular-momentum}
  p_\varphi' = p_\varphi.
\end{equation}
Given the transformations in
\cref{eq:rotating-scalar-potential,eq:rotating-flux-function,%
eq:rotating-angular-momentum}, the effective
potential transforms according to
\begin{equation}
  \label{eq:rotating-potential}
  U' = U - \Omega p_\varphi.
\end{equation}
This is obviously consistent with \cref{eq:rotating-angular-velocity} since
$\omega=\pader{U}{p_\varphi}$. From \cref{eq:rotating-potential} it follows
immediately that
\begin{equation}
  \label{eq:rotating-variation}
  \delta U' = \delta U
  \quad\textrm{and}\quad
  \delta^2 U' = \delta^2 U
\end{equation}
for arbitrary variations $\delta\cyl$ and $\delta z$. This means that neither
the location of circular orbits nor their energetic stability is affected by a
transformation to a rotating frame of reference.

In the rotating frame, the components of the poloidal electric field are given
by
\begin{equation}
  \label{eq:E-rho-prime}
  \begin{aligned}
    E_\cyl' &= E_\cyl + \cyl\Omega B_z + \cyl\Omega^2 \\
    E_z' &= E_z - \cyl\Omega B_\cyl.
  \end{aligned}
\end{equation}
The second term on each right hand side arise simply from a Galilean
transformation with relative velocity $\cyl^2\Omega\nabla\varphi$. The last
term in \cref{eq:E-rho-prime} is the centrifugal force. The components of the
transformed poloidal magnetic field are
\begin{equation}
  \label{eq:rotating-magnetic-field}
  \begin{aligned}
    B_\cyl' &= B_\cyl \\
    B_z' &= B_z + 2\Omega,
  \end{aligned}
\end{equation}
which include the Coriolis force. The toroidal component of the magnetic field
$B_\varphi$ is invariant. In light of
\cref{eq:dispersion-relation,eq:rotating-variation} this means that the
spectral stability of circular orbits is not affected either by a
transformation to rotating frame of reference.

From Gauss' law \labelcref{eq:gauss} and Ampère's law \labelcref{eq:ampere} it
follows that
\begin{equation}
  \label{eq:rotating-charge}
  {\charge}' = \charge - \cyl\Omega\current + 2\Omega B_z + 2\Omega^2
\end{equation}
and ${\current}'=\current$. It is easy to check that these transformations
together with \cref{eq:rotating-angular-velocity,eq:rotating-magnetic-field}
indeed leave the trace of the Hessian as given in \cref{eq:trace} invariant.

\Cref{eq:rotating-charge} shows that regions of space that are source-free in
an inertial frame are not source-free in a rotating frame. We stress,
however, that the last term in \cref{eq:rotating-charge} is not a physical
source of either the electromagnetic or gravitational fields, but is a
fictitious charge density that derives from the centrifugal potential via
Gauss' law \labelcref{eq:gauss}.

\begin{acknowledgments}
  We thank Pablo Benítez-Llambay, Luis García-Naranjo and Jihad Touma for
  insightful comments. We are grateful for the hospitality of the Institute
  for Advanced Study where part of this work was carried out. The research
  leading to these results has received funding from the European Research
  Council (ERC) under the European Union's Seventh Framework programme
  (FP/2007--2013) under ERC grant agreement No 306614.
\end{acknowledgments}


\end{document}